\documentclass[letterpaper]{emulateapj}

\usepackage{amsmath}
\usepackage{amssymb}
\usepackage{xspace}
\usepackage{color}
\usepackage{ifthen}
\usepackage{xspace}

\newcommand{\powersep}{{\ensuremath{\times}}}

\newcommand{\mn}{\mu_\nu}
\newcommand{\m}{\mu_{12}}
\newcommand{\muB}{\mu_{\mathrm{B}}}
\newcommand{\me}{m_{\mathrm{e}}}

\newcommand{\sP}{\sigma^{\mathrm{pair}}}
\newcommand{\QP}{Q^{\mathrm{pair}}}
\newcommand{\eP}{\varepsilon_{\mathrm{pair}}}
\newcommand{\ePl}{\varepsilon_{\mathrm{plasma}}}
\newcommand{\ePh}{\varepsilon_{\mathrm{photo}}}
\newcommand{\Qu}{\QP_{\mu}}

\newcommand{\Qsm}{\QP_{\mathrm{SM}}}

\newcommand{\vrel}{v_{\mathrm{rel}}}
\newcommand{\gad}{\gamma_{\mathrm{ad}}}
\newcommand{\GF}{G_{\mathrm{F}}}
\newcommand{\CV}{C_{\mathrm{V}}}
\newcommand{\CA}{C_{\mathrm{A}}}
\newcommand{\tW}{\theta_{\mathrm{W}}}

\newcommand{\opl}{\omega_{\rm pl}}
\newcommand{\ov}{\overline}
\renewcommand{\to}{\rightarrow}
\newcommand{\K}{{\ensuremath{\mathrm{K}}}\xspace}

\newcommand{\Msun}{{\ensuremath{\mathrm{M}_{\odot}}}\xspace}

\newcommand{\erggs}{{\ensuremath{\mathrm{erg}\,\mathrm{g}^{-1}\,\mathrm{s}^{-1}}}\xspace}

\newcommand{\gcc}{{\ensuremath{\mathrm{g}\,\mathrm{cm}^{-3}}}\xspace}

\newcommand{\eV}{{\ensuremath{\mathrm{e}\!\mathrm{V}}}\xspace}
\newcommand{\keV}{{\ensuremath{\mathrm{k}\eV}}\xspace}

\newcommand{\GeV}{{\ensuremath{\mathrm{G}\eV}}\xspace}
\newcommand{\TeV}{{\ensuremath{\mathrm{T}\eV}}\xspace}

\newcommand{\lSect}[1]{{\label{sec:#1}}}
\newcommand{\lFig}[1]{{\label{fig:#1}}}

\newcommand{\lTab}[1]{{\label{tab:#1}}}

\newcommand{\Tabff}[1]{{\ref{tab:#1}}}
\newcommand{\Tab}[1]{{Table~\Tabff{#1}}}

\newcommand{\PAN}[1]{{{#1}}}

\newcommand{\FIGFF}[2]{{\ref{fig:#2}\PAN{#1}}}

\newcommand{\FIG}[2]{{Fig.~\FIGFF{#1}{#2}}}
\newcommand{\Fig}[1]{{\FIG{}{#1}}}

\newcommand{\Sectff}[1]{{\ref{sec:#1}}}

\newcommand{\Appendix}[1]{{Appendix~\Sectff{#1}}}

\newcommand{\isofont}[1]{{\mathrm{#1}}}
\newcommand{\isomass}[1]{{\ensuremath{\isofont{^{#1}}}}}
\newcommand{\isocharge}[1]{{\ensuremath{\isofont{_{#1}}}}}
\newcommand{\isotope}[3]{{\ensuremath{\isocharge{#1}\isomass{#2}\isofont{#3}}}}

\newcommand{\I}[2]{{\isotope{}{#1}{#2}}}
\newcommand{\El}[1]{{\I{}{#1}}}

\newcommand{\Ep}[1]{{\ensuremath{10^{#1}}}}

\newcommand{\E}[1]{{\ensuremath{\powersep\Ep{#1}}}}

\newcommand{\Ye}{{\ensuremath{Y_{\mathrm{e}}}}\xspace}

\newcommand{\Tc}{{\ensuremath{T_{\mathrm{c}}}}\xspace}
\newcommand{\rhoc}{{\ensuremath{\rho_{\mathrm{c}}}}\xspace}

\newcommand{\nue}{{\ensuremath{\nu_{\El{e}}}}\xspace}

\slugcomment{Draft for ApJ, October 16, 2008}

\shorttitle{Neutrino Magnetic Moment in Massive Stars}
\shortauthors{Heger, Friedland, Giannotti, \& Cirigliano}

\begin{document}

\title{The Impact of Neutrino Magnetic Moments on the Evolution of Massive Stars}

\author{Alexander~Heger}
\affil{Theoretical
School of Physics and Astronomy, 
University of Minnesota, 
116 Church ST SE,
Minneapolis, MN 55455\\
Astrophysics Group, MS B227, Los Alamos National Laboratory, Los
Alamos, NM 87545;\\
Department of Astronomy and Astrophysics,
University of California, Santa Cruz, CA 95064}
\email{alex@physics.umn.edu}

\author{Alexander~Friedland and Maurizio Giannotti}
\affil{Elementary Particles and Field Theory Group, MS B285, Los Alamos National Laboratory, Los Alamos, NM 87545}
\email{friedland@lanl.gov}
\email{maurizio@lanl.gov}


\author{Vincenzo~Cirigliano}
\affil{Nuclear Physics Group, MS B283, Los Alamos National Laboratory, Los
Alamos, NM 87545}
\email{cirigliano@lanl.gov}



\begin{abstract}
  
  We explore the sensitivity of massive stars to neutrino magnetic
  moments.  We find that the additional cooling due to the neutrino
  magnetic moments bring about qualitative changes to the structure
  and evolution of stars in the mass window $7\,\Msun \lesssim M
  \lesssim 18\,\Msun$, rather than simply changing the time scales for
  the burning. We describe some of the consequences of this modified
  evolution: the shifts in the threshold masses for creating
  core-collapse supernovae and oxygen-neon-magnesium white dwarfs and
  the appearance of a new type of supernova in which a partial
  carbon-oxygen core explodes within a massive star.  The resulting
  sensitivity to the magnetic moment is at the level of $(2-4) \times
  10^{-11}\,\muB$.

\end{abstract}

\keywords{neutrinos --- stars: supergiants --- stars: evolution
  --- stars: interiors}

\section{Introduction}

Cooling by neutrino emission is known to be important in a variety of
stellar systems, from low-mass red giants (RG) and horizontal branch
(HB) stars to white dwarfs (WD), neutron stars and core-collapse
supernovae (CCSN). Although, compared with photons, neutrinos interact
extremely weakly, once produced they easily escape from stellar
interiors carrying away energy that would otherwise take much longer
to be transported to the surface by radiation or convection. The
resulting energy sink in the center of the star can dictate the star's
rate of nuclear burning, structure and evolution, and ultimately how
it ends its life.  If a new interaction modifies the rate at which
neutrinos are produced, the evolution of the star could change. Stars
can therefore be used as laboratories to search for non-standard
properties of the neutrino \citep{Bernstein}.

One such new interaction that will be considered in this paper are the
neutrino (transition) magnetic moments.  
The magnetic moments couple neutrinos to photon through
the effective Lagrangian term
\begin{equation}
\label{Lint}
L=-\frac12 \mu_\nu^{ij} \, \overline \psi_i \sigma_{\alpha\beta}\psi_j\, F^{\alpha\beta} \,,
\end{equation}
where $\psi$ is the neutrino field, $F$ the electromagnetic field
tensor, $\alpha$, $\beta$ are Lorentz indices, $i$, $j$ are the flavor
indices. This interaction is non-zero, although extremely small,
already in the Standard Model (SM) with non-zero neutrino mass.  In some
extensions of the SM it is allowed to be as large as the
present direct laboratory bound of $\mu_\nu \lesssim 10^{-10}\,\muB$
\citep{MUNU}, where $\muB=e/2\me$ is the Bohr magneton
%
%
(see Sect.~\ref{sect:operatoranalysis}).  Such new physics could
reside just above the electroweak scale of $100\,\GeV$.

The effects of the magnetic moment interaction on the evolution of
low-mass stars have been extensively studied, as will be reviewed in
Sect.~\ref{sect:munu_review} (see also, e.g.,
\citealt{RaffBook,RaffRep99,Raff00} for a review).  In fact, these
stars are thought to give the best  bound on the neutrino
magnetic moment, about an order of magnitude more restrictive than the
direct laboratory bound.  The effect of the magnetic moment on the
evolution of massive ($M \gtrsim 7\, \Msun$) stars, however, has not
been established.
This paper is an exploratory investigation designed to fill this gap. Our
goal is to answer the following questions:
\begin{itemize}
\item Does the presence of a non-vanishing neutrino magnetic moment
  influence in any way the behavior of a massive star?
\item If so, for which range of stellar masses?
\item What are the induced effects?
\item What is the sensitivity of massive stars to the size of the
  neutrino magnetic moment?
\end{itemize}

The paper is organized as follows: In Sect.~\ref{sect:munu_review}, we
discuss the neutrino magnetic moments from the theoretical and
phenomenological points of view. In particular, we review the present
bounds from experiments and astrophysical considerations.  In
Sect.~\ref{sect:analytical}, we review the standard evolution of
massive stars and consider how might change with the addition of the
extra cooling induced by a non-zero $\mn$. In
Sect.~\ref{sect:numerical}, we present the results of our numerical
modeling. Finally in Sect.~\ref{sect:conclusions}, we summarize our
results and discuss possible future directions.

\section{Neutrino Magnetic Moment}
\label{sect:munu_review}

\subsection{Theoretical considerations}
\label{sect:operatoranalysis}
        
The search for the neutrino magnetic moment goes back to
the neutrino discovery experiment by~\citet{Cowan}, who put the first
bound on the magnetic moment, $\mu_\nu\lesssim 10^{-9}\,\muB$ 
%

On the theoretical front, it was realized many decades ago that a
neutrino with a non-zero mass should have a non-vanishing magnetic
moment even with the Standard Model interactions~\citep{marciano,
  lee,fujikawa}. Due to the specific nature of the SM interactions,
however, namely, coupling of $W$ to left-handed currents only, the
induced value of the magnetic moment turns out to be very small,
$\mu_\nu/\,\muB \simeq 3 \times 10^{-19} \times (m_\nu/1\,\eV)$. In
extension of the SM, this limitation need not apply and much larger
values of $\mu_\nu$ can be obtained.

There are several classes of models beyond the SM predicting a large
magnetic moment (as large as $\mu_\nu/\muB \sim 10^{-10}$).  These
scenarios typically manage to simultaneously have a small neutrino
mass and a large magnetic moment by one of the two following
mechanisms: (i) requiring that the new physics dynamics obeys some
(approximate) symmetry that forces $m_\nu=0$ while allowing
non-vanishing
$\mu_\nu$~\citep{Voloshin,barbieri-mo,Georgi,mohapatra1,Grimus}; or by
(ii) engineering a spin suppression mechanism that keeps $m_\nu$
small~\citep{Barr}.

It is important to stress that the value of the magnetic moment in SM
extensions cannot be arbitrarily large. 
It can be shown in a model-independent way that a large value of 
$\mu_\nu$ would induce, via electroweak
radiative corrections, a neutrino mass  above the
current experimental limit
(conservatively, $\sim 1\,\eV$).  Quite interestingly, the resulting
\lq\lq naturalness" upper limits
are substantially different for Dirac and Majorana neutrinos, due to
the different nature and breaking of the approximate symmetries
imposed to simultaneously ensure $m_\nu \simeq 0$ and $ \mu_\nu \neq
0$.
For the Dirac case, a general effective theory analysis leads to
\citep{Bar-Fio,dirac}:
\begin{equation}
\frac{\mu_\nu}{\muB} \ \alt 
\  3  \times 10^{-15} \times \left(   
\frac{\delta m_\nu}{1\,\eV}   
\right) \, \left( \frac{1\,\TeV}{\Lambda} \right)^2~, 
\label{eq:nat-dirac}
\end{equation} 
where $\delta m_\nu$ is the induced mass term and $\Lambda$ is the
(unknown) new physics mass scale at which the magnetic moment
originates.
For the Majorana transition magnetic moments one finds
\citep{Davidson,majorana}:
\begin{equation}
\small
\frac{\mu_\nu^{\alpha\beta}}{\muB} \ \alt \ 4 \times 10^{-9}
\left(\frac{\left[\delta m_\nu\right]_{\alpha\beta}}{1\,\eV}\right)
\left(\frac{1\,\TeV}{\Lambda}\right)^2 
\left| \frac{m_\tau^2}{m_\alpha^2 - m_\beta^2} \right|~
\label{eq:nat-majorana}
\end{equation}
where $(\delta m_\nu )_{\alpha \beta}$ is the induced contribution to
the neutrino mass matrix and $m_{\alpha}, m_{\beta}$ represent the \emph{
  charged lepton} masses for the flavors $\alpha,\beta$.
Therefore, the naturalness limit on the magnetic moment of Majorana
neutrinos is considerably weaker.
Note that in both cases there is a strong quadratic dependence on the
new physics scale $\Lambda$, for which we have taken the reference
value of $\Lambda = 1\,\TeV$.

In summary, evidence of $\mu_\nu > 10^{-19}\,\muB$ would indicate the
need for a radical extension of the weak sector of the SM.
Moreover,
evidence of a magnetic moment in the window $10^{-15}\,\muB \alt \mn \alt
10^{-10}\,\muB$ would provide strong evidence for the
Majorana nature of neutrinos. Therefore, astrophysical probes of the
neutrino magnetic moment  with sensitivity right below 
$\mu_\nu \sim 10^{-10}\,\muB$ are of considerable theoretical value.

\subsection{Current Bounds}

The first experimental bound on the magnetic moment of the electron
neutrino, $\mu_\nu\lesssim 10^{-9}\,\muB$, was obtained by
\citet{Cowan}, by analyzing electron recoil spectra in (anti)-neutrino
electron scattering.  The same technique has led, over time, to more
stringent bounds, $\mn< (5-10)\times 10^{-11}\,\muB$, at $90\,\%$ C.L.
\citep{MUNU,LastLimit}.  A similar bound ($\mn< 5.4\times
10^{-11}\,\muB$, at $90\,\%$ C.L.)  was recently confirmed by the
Borexino collaboration \citep{Borexino} from the analysis of solar
neutrinos.

Another set of constraints comes from the analysis of energy loss in stars.
This method was pioneered by \citet{Bernstein} who considered the
effect of the addition cooling due to the neutrino magnetic moment on
the lifetime of the Sun and estimated that $\mn \lesssim
10^{-10}\,\muB$. A recent bound from the Sun is $\mu_\nu\lesssim
4\times 10^{-10} \,\muB$ \citep{RaffRep99}.  A more stringent
bound, $\mn\lesssim3 \times 10^{-12}\,\muB$, comes from the analysis of
the RG branch, where the extra cooling could delay the
onset of the helium flash
\citep{Raff1,Raff2,Raff3,Castellani,Catelan}.  A comparable bound also
follows from the analysis of WD cooling \citep{Blinnikov}. In
contrast, the effect of a non-zero neutrino magnetic moment on the
cooling of a neutron star, studied by \citet{Iwamoto}, leads to the
considerably less stringent constraint $\mu_\nu\lesssim 5\times
10^{-7} \,\muB$, though a value as small as $\sim 10^{-10} \,\muB$ would
change the cooling of the crust and could be observable in the future.
Finally, HB stars are also quite sensitive to a
non-vanishing magnetic moment. A value of $\mu_\nu$ of about
$1.5\times 10^{-11} \,\muB$ would shorten the lifetime of the He-burning
stage of those stars, and consequently change the number ratio of HB
stars versus RG stars in globular clusters above the observational
bound \citep{RaffHB,RaffBook}.
An important feature of these constraints is that they do not depend
on the nature -- Dirac or Majorana -- or flavor of the neutrinos.

\begin{figure}
\centering
\includegraphics[angle=0, width=\columnwidth]{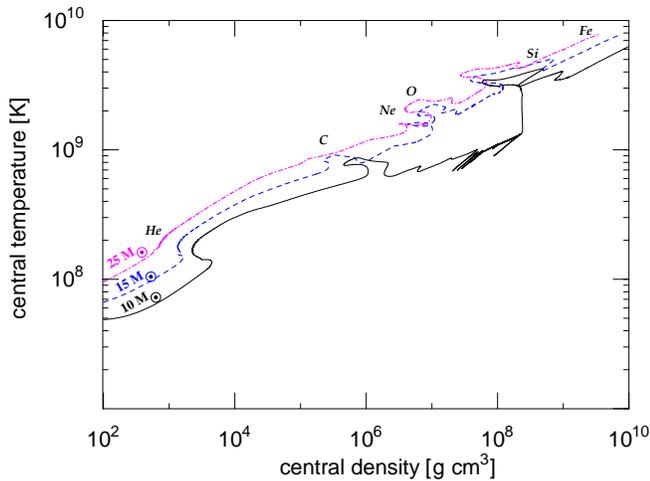}
\caption{Standard model evolution of the central temperature and
  density for a $25\,\Msun$, $15\,\Msun$ and $10\,\Msun$ star.  The
  different burning stages are indicated by the corresponding chemical
  symbol.}
 \label{first}
\end{figure}

For Dirac neutrinos with a non-vanishing magnetic moment, the left
handed states could be transformed into right handed states in an
external electromagnetic field. For a supernova (SN), where left
handed neutrinos are trapped and right handed neutrinos are not, this
would mean a very efficient cooling mechanism. This leads to the bound
$\mu_\nu\lesssim 3\times 10^{-12} \,\muB$~\citep{SN,lattimer,ayalaetal}.
A comparable bound comes from the observation that some of the right
handed neutrinos could transform back into the active ones in the
galactic magnetic field, and should have been observed by the
detectors which measured the neutrino signal from
SN~1987A\citep{Notzold}.  

Similarly, the analysis of big bang nucleosynthesis puts
limits on the magnetic moment for Dirac neutrinos. A large magnetic
moment would bring right handed neutrinos in thermal equilibrium
before the cosmological nucleosynthesis, spoiling the 
prediction of light elements abundances.  The current bound is
approximately $\mu_\nu\lesssim 10^{-11} \,\muB$ (see \citealt{Dolgov}
and references therein).

Lastly, a Majorana neutrino with a non-vanishing magnetic moment could
be transformed into an anti-neutrino of a different flavor in an
external electromagnetic field.  \citet{Valle04} inferred the bound of
a few $10^{-12} \,\muB$ from the lack of the observation by KamLAND of
the antineutrino flux from the Sun. Using the same data,
\citet{Friedland2005}, however, argued that the bound is actually an
order of magnitude weaker.

To conclude this Section, we point out that the bounds discussed above
are sensitive to different effective magnetic moments, namely to
different combinations of the elements of the magnetic moment matrix
$\mn^{\alpha \beta}$.  The terrestrial experiments using reactor
sources (electron anti-neutrinos) are sensitive to $(\mu_\nu)_{e}^2 =
\sum_{\alpha} | \mn^{e \alpha}|^2$.  On the other hand, the bounds
based on solar neutrinos analyses are sensitive to a weighted average
of $(\mu_\nu)_{e}^2$, $(\mu_\nu)_{\mu}^2$, $(\mu_\nu)_{\tau}^2$, with
roughly equal weights determined by the oscillation probabilities.
Finally, the bounds based on energy loss in stars are sensitive to
$\mu_\nu^2 = \sum_{\alpha,\beta} | \mn^{\alpha \beta}|^2$.

\section{Massive Stars and non-standard Cooling: Analytical Considerations}
\label{sect:analytical}

\subsection{Review of the standard evolution and cooling}

\begin{figure}
\centering \includegraphics[angle=0,
  width=\columnwidth]{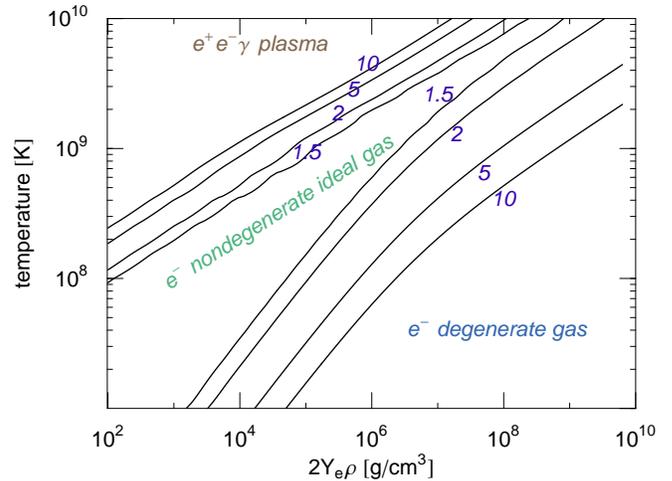}
\caption{The contours of the ratio $P_{\mathrm{e}+\gamma}/P_\mathrm{e,
    i.g.}$ of the combined electron and photon pressure
  $P_{\mathrm{e}+\gamma}$ to the corresponding electron ideal gas
  pressure, $P_{e, i.g.}\equiv n_\mathrm{e} kT$, as a function of
  temperature and density of the medium.}
 \label{fig:pressure}
\end{figure}

In this Subsection, for the convenience of the reader, we briefly
review the main features of the standard evolution of massive stars
relevant to our subsequent discussion of the effects of the neutrino
magnetic moment.  We collect some of the technical details in Appendix
A.  For a standard introduction to the evolution of massive stars,
see, e.g., \citet{Clayton, Arnett, Heger}.

We consider here stars in the mass range of $7\, \Msun \lesssim M
\lesssim 25\, \Msun$.  The basic information about the physical
conditions in the core of these stars can be obtained from simple
considerations of hydrostatic equilibrium and balance between energy
generation and loss. Assuming for the moment that the interiors of
these stars can be described as ideal, nondegenerate gas, from the
consideration of hydrostatic equilibrium one can find a simple
relation among the mass $M$, central temperature $\Tc$ and central
density $\rho_\mathrm{c} $ \citep{Clayton,Arnett}:
\begin{equation}
\label{core}
\left(\frac{\Tc}{1\,\K}\right) \simeq 4.6\times 10^6 \, \mu\, 
\left(\frac{\! M}{\, \Msun}\right)^{2/3}\,\left(\frac{\rhoc}{1\,\gcc}\right)^{1/3} \,,
\end{equation}
where $\mu=(\sum_i Y_i+\Ye)^{-1}$, with $\Ye$ the mean number of
electrons per baryon, and $Y_i=X_i/A_i$ with $X_i$ and $A_i$ the mass
fraction and atomic number of species $i$ respectively.  

The curves in Fig.~\ref{first} represent the evolution of the central
temperature and density for stars of three different masses, as found
with our numerical code. 
Once one fuel is exhausted, the star
follows a period of compression, which lasts until the temperature and
density satisfy the required condition for the ignition of another
nuclear reaction.
We can see that these stars indeed
approximately follow Eq.~(\ref{core}): in particular, the more massive
ones have higher central temperatures. This simple fact will have
important implications later.

The approximation of ideal nondegenerate gas is valid for a relatively
limited set of densities and temperatures, as seen in
Fig.~\ref{fig:pressure} which shows the ratio of the combined electron
and photon pressure $P_\mathrm{e}+P_\mathrm{rad}\equiv
P_{\mathrm{e}+\gamma}$ to the corresponding electron ideal gas
pressure, $P_\mathrm{e, i.g.}\equiv n_\mathrm{e}kT$.  At high
temperatures and low densities the pressure is dominated by radiation
(photons and relativistic electron-positron plasma).  At lower
temperatures and high densities, the pressure is set by electron
degeneracy.

Comparing Figs.~\ref{first} and \ref{fig:pressure}, we see that the
evolutionary trajectory of the $25\, \Msun$ star indeed lies almost
entirely in the region where the ideal gas equation of state applies,
except at the very last stages of the evolution which are on the edge
of being degenerate.  The lower mass stars considered in
Fig.~\ref{first}, having lower central temperatures, are more subject
to the effects of degeneracy which set in after the start of the
carbon burning \citep{Arnett}.

Two important observations need to be made at this point. First, so
long as there is steady nuclear burning at the center, the star has to
be outside, or at least on the edge, of the degenerate region.
Indeed, a strongly degenerate system does not have the necessary
negative feedback mechanism to regulate the rate of burning: energy
production does not lead to expansion and cooling of the central
region if the pressure is independent of temperature, as is the case
for strongly degenerate systems (see, e.g., \citealt{RaffBook}).  The
detour of the $10\,\Msun$ star through the region of high degeneracy
seen in Fig.~\ref{first} indicates that the nuclear burning at this
stage happens \emph{in a shell} rather than in the core \citep{Heger}.
When the burning reaches the center, the temperature jumps and the
feedback is enabled. As we will see, this behavior will be even more
pronounced in the presence of the neutrino magnetic moment. 

Second, there is an important physical distinction between the regimes
of non-relativistic and relativistic degeneracy. The former is
characterized by an adiabatic exponent $\gad=5/3$ (polytropic index
$n=3/2$)\footnote[$\dagger$]{We use the standard definition
  $P=K\rho^{\gad}=K\rho^{1+1/n}$ relating density $\rho$ and pressure
  $P$.} while the latter has $\gad=4/3$ ($n=3$). A non-relativistic
degenerate gas responds more strongly to compression than a
relativistic one. In fact, $\gad=4/3$ corresponds to the edge of
stability: if a sufficient fraction of the stellar core has this
adiabatic exponent, the star becomes unstable to collapse
\citep{ZeldovichBook}. In the case of white dwarfs, this leads to the
well-known upper limit on the mass, the Chandrasekhar mass.  In our
case of extra cooling, this will lead to the appearance of a new type
of supernova.

We now turn to the mechanisms of energy loss. Energy generated by
nuclear reactions is lost either through the surface of the star as
photon radiation, or through the volume via neutrino emission. Photon
energy loss per unit mass is about $10^4-10^5\,\erggs$ for stellar
masses considered here (see, e.g., \citealt{Arnett}, \S~6.4). Neutrino
cooling becomes important when the emission rate per unit mass exceeds
this value.

At the temperatures and densities of interest for our discussion,
neutrinos are dominantly produced by the following processes (see
\citealt{Salpeter, Dicus} and reference therein): 
\begin{enumerate}
\item \emph{photo process},
$\gamma\, e^-\to e \, \nu\, \ov\nu$; 
\item
\emph{pair process}, $e^+\,e^- \to \nu\, \ov \nu$;
\item
\emph{plasma process}, $\gamma\to \nu\, \ov\nu$,\\ also known as plasmon
decay; 
\item 
\emph{bremsstrahlung}, $e^-(Ze)\to (Ze)\,e^-\, \nu\, \ov\nu$. 
\end{enumerate}
The diagrams for the first three of these processes are depicted in
Fig.~\ref{diagramSTD}. 

\begin{figure}
  \centering \includegraphics[angle=0,
  width=0.5\textwidth]{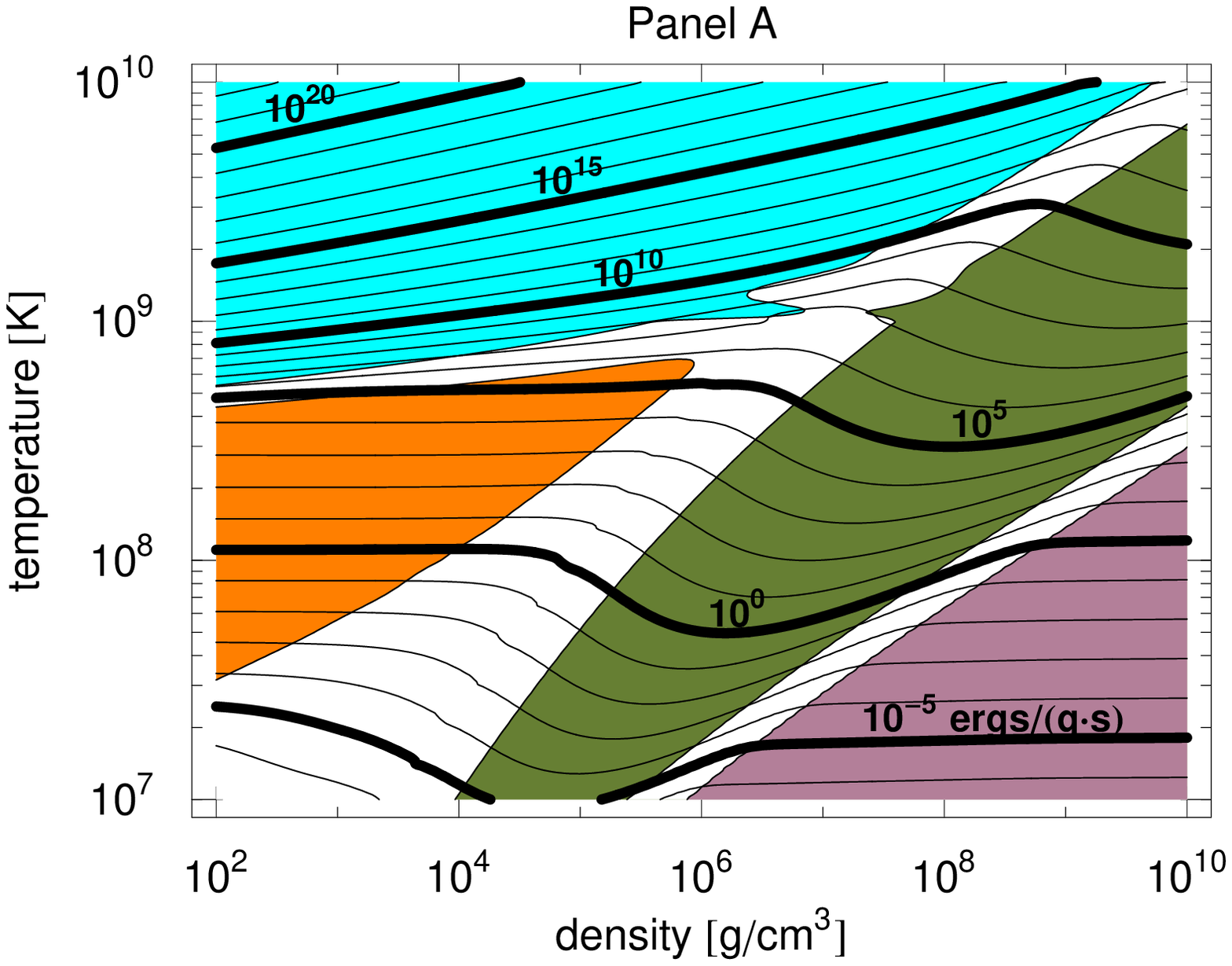}
  \includegraphics[angle=0,
  width=0.5\textwidth]{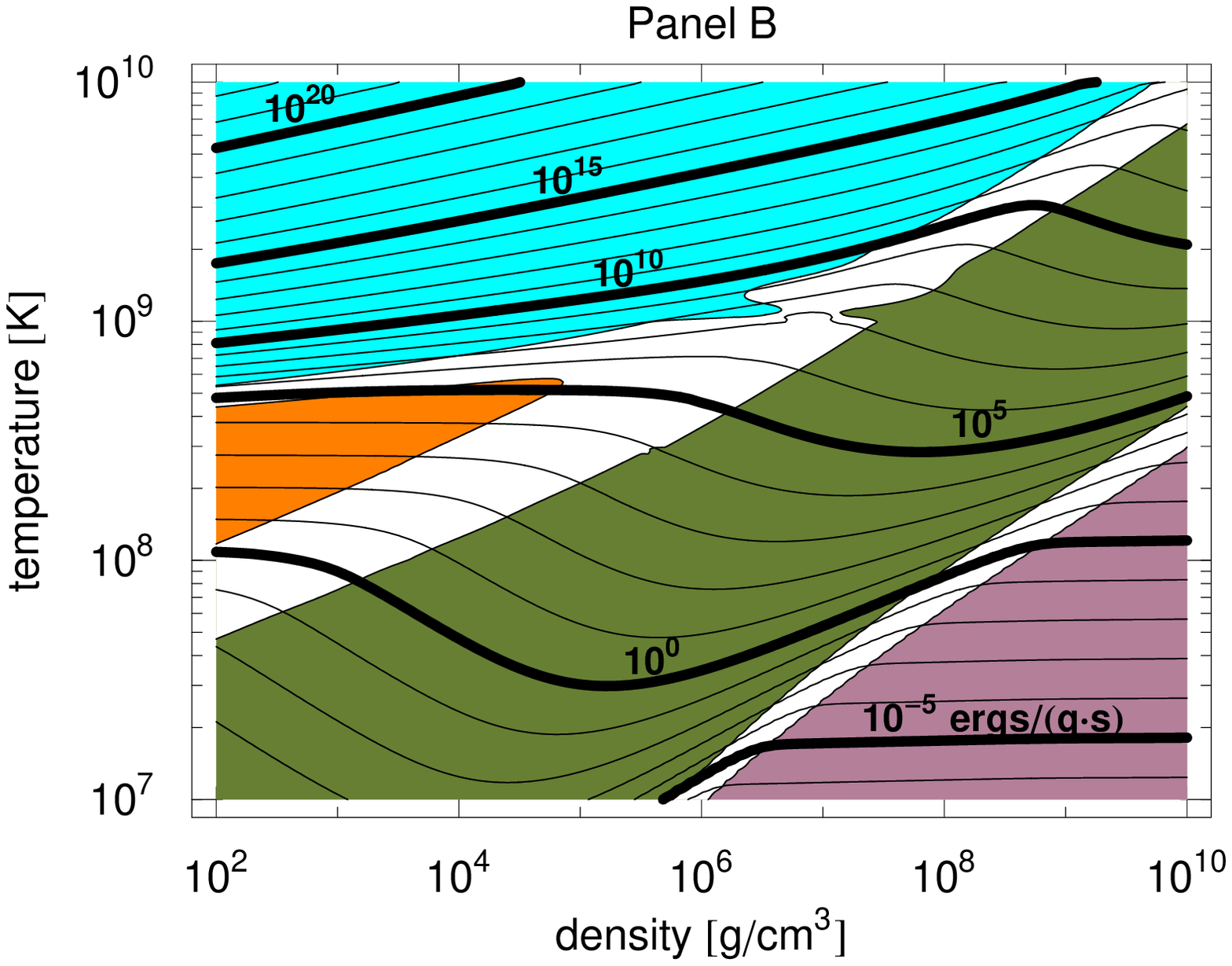}
  \includegraphics[angle=0,
  width=0.5\textwidth]{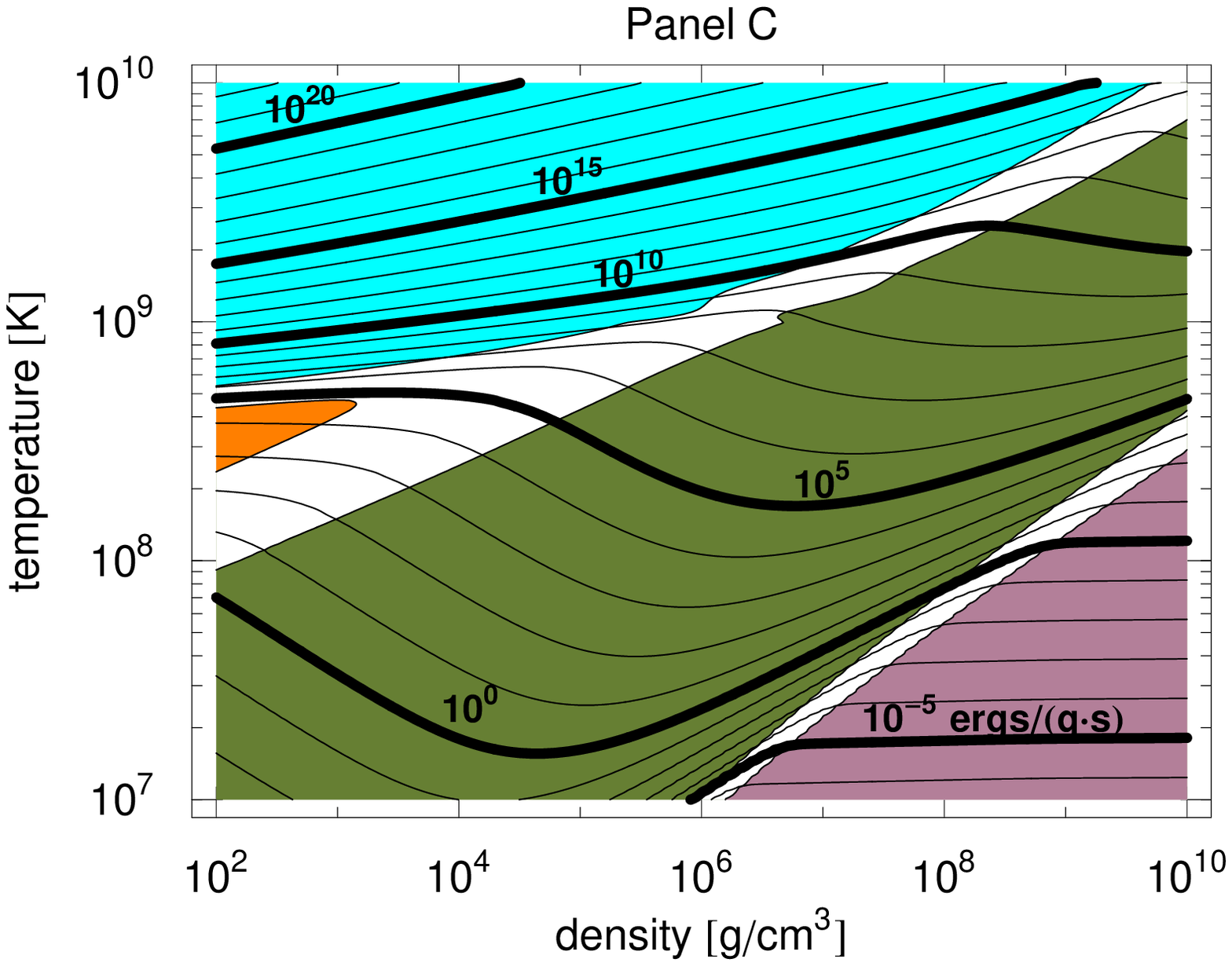}
\caption{Neutrino energy loss. From above
{\bf Panel~a (top):} Standard energy loss, $\mn=0$;
{\bf Panel~b (middle):} Energy loss for  $\mn=10^{-11}\,\muB$;
{\bf Panel~c (bottom):} Energy loss for  $\mn=5\times 10^{-11}\,\muB$.
The different colors show the regions where the relevant process is
$90\,\%$ or more of the total neutrino cooling: Turquoise (top)
neutrino pair production $e^+\,e^- \to \nu\, \ov \nu$; Orange (middle
left) neutrino photo production $\gamma\, e^-\to e \, \nu\, \ov\nu$;
Green (middle low) neutrino production by plasmon decay $\gamma\to
\nu\, \ov\nu$; Purple (bottom right corner) bremsstrahlung production
$e^-(Ze)\to (Ze)\,e^-\, \nu\, \ov\nu$.  For comparison, see Fig.~C.4
in \citet{RaffBook}.
}
 \label{second}
\end{figure}

\begin{figure}
\centering
\includegraphics[angle=0, width=0.25\textwidth,clip=true]{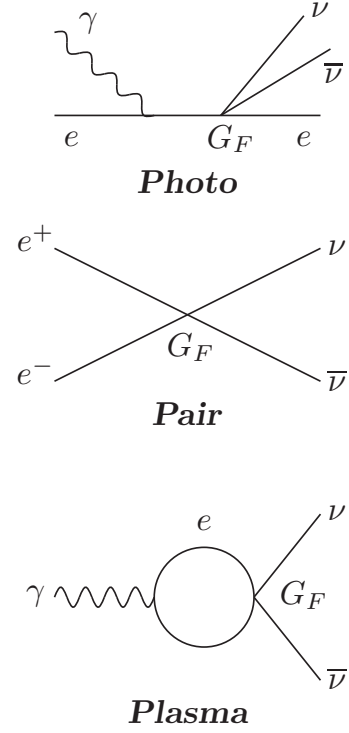}
\caption{Feynman diagrams for the standard model neutrino production processes
relevant for our discussion. From top to bottom:
photo process, $\gamma\, e^-\to e \, \nu\, \ov\nu$;
pair process, $e^+\,e^- \to \nu\, \ov \nu$;
plasma process, $\gamma\to \nu\, \ov\nu$.
}
 \label{diagramSTD}
\end{figure}

Which of these processes dominates depends on the density and
temperature, as seen in Fig.~\ref{second}a.  The contour lines show
the rate of energy loss, and are computed using the expressions for
the neutrino cooling rates given in~\citet{Itoh}.  These rates are
used as a starting point in our numerical analyses, to compute the
reference models with standard particle physics.

Superimposing Fig.~\ref{first} to Fig.~\ref{second}a, we see that in
the early burning stages (H- and He-burning), the neutrino production
is dominated by the photo process.  The energy loss (see the
Eq.~(\ref{photo}) or contour lines in Fig.~\ref{second}a), however, is
still small compared to the stellar luminosity, $10^4-10^5\,\erggs$.
The neutrino energy loss starts dominating over the radiation only
after central helium depletion, for temperatures above $\sim 5\times
10^8\,\K$.  When carbon is ignited, $T\simeq 7\times 10^8\,\K$, the
star is already losing most of its energy through neutrinos.  For the
more massive stars (see the curve corresponding to $25\,\Msun$ and
$15\,\Msun$ in Fig.~\ref{first}), neutrinos will be mainly produced by
the pair process.  For lower mass stars, however, there will be a
competition between the pair and plasma processes.  As we see the
bremsstrahlung process is never important for a massive star.

Since the neutrino emission rate increases rapidly with
temperature (contours lines in Fig.~\ref{second}a), the fuel is
consumed faster in the late stages of stellar evolution.
For this reason, these later
stages are characterized by progressively shorter time scales.  In
fact, a massive star spends about $90\,\%$ of its life burning H, and
almost all of the remaining $10\,\%$ burning He.  The time scale for
the other burning stages is thousands of years for C, years for Ne and
O, and only days for Si \citep{Heger}.

In our
numerical model (see details in Sect.~\ref{sect:numerical}), stars
whose initial mass is above $\sim 9\,\Msun$ can ignite all burning
phases (H, He, C, Ne, O, and Si burning)\footnote{For a
discussion of how this threshold value varies between different codes,
see \citep{Heger2}}.  They end up with a typical ``onion structure''
with iron in the center and several layers around, in which fusion of
lighter elements occurs.  No energy can be gained by fusion reactions
in the iron core, and once its mass exceeds $\sim 1.4\,\Msun$
the star collapses producing a CCSN.  

Stars less massive than about $9\, \Msun$ have a different destiny.  If
the initial mass is less than about $7.5\, \Msun$ they never reach the
stage of carbon-burning and end their life as carbon-oxygen white
dwarfs (CO-WDs, see the bottom row of Fig.~\ref{final}).  If their
initial mass is in the range $7.5 \lesssim M/\Msun \lesssim 9$, they
do ignite carbon but not neon, and end their lives as
oxygen-neon-magnesium white dwarfs (ONeMg WD). These numbers are
important to us because they will be shifted by the presence of the
additional cooling.  Moreover, a new evolution possibility will be
opened up.

We also note that the stars that we just classified as making ONeMg
WDs could yet have another fate.  They are close to the Chandrasekhar
mass, and, after a typical ``deep third dredge-up'' that mixes the
envelope with helium layer of the core, the core may continue to grow
by hydrogen/helium shell burning. Eventually, collapse of the core may
occur due to electron capture and a supernova can result (electron
capture supernova, ECSN, \citealt{Miyaji1980}).  Such a supernova
typically would produce only very little \I{56}{Ni}.  The details of
whether the core can grow highly depends on the dredge-up (mixing)
after and between the thermal pulses of the hydrogen/helium shell
burning, competing with the mass loss of star that terminates core
growth when the envelope has been lost. The historical developments on
this topic are summarized in \citet{Heger}, while some recent
developments are discussed in, e.g., \citet{Heger2}.  Due to the uncertainty of the
mass range of ECSN, we will consider them as part of our ONeMg core
``bin'' and just note that they comprise some fraction of the upper
end of the ONeMg WD mass range.

Also note that we only consider single star evolution here.  Special
evolution paths that may occur in binary star systems, like, e.g.,
Type I SNe, are not considered in this paper.

\subsection{Effects of the $\mu_\nu$-induced  extra cooling}

\begin{figure}
\centering
\includegraphics[clip=true,angle=0, width=0.25\textwidth]{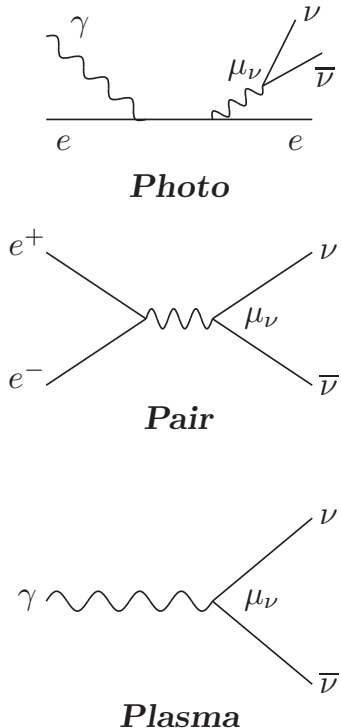}
\caption{Feynman diagrams for the electromagnetic contribution
to the neutrino production processes relevant for this paper.
From top to bottom:
photo process, $\gamma\, e^-\to e \, \nu\, \ov\nu$;
pair process, $e^+\,e^- \to \nu\, \ov \nu$;
plasma process, $\gamma\to \nu\, \ov\nu$.
}
 \label{diagramMu}
\end{figure}

If neutrinos have a non-vanishing magnetic moment, additional
processes shown in Fig.~\ref{diagramMu} must be considered.  It turns
out that the main effect of the neutrino magnetic moment is to
increase the plasmon decay rate. For example, for $\mn/\muB =
10^{-10}$, density $\rho \sim 10^4\,\gcc$ and temperature $T =
10^8\,\K$, the non-standard plasma process energy loss is about $10^3$
larger than the standard one (see Eq.~\ref{plasma}).  On the other
hand, for the same value of $\mn$ the non-standard pair production
rate never exceeds $30\,\%$ of the standard one (see
Fig.~\ref{fig:Qpair}).  Details of the analysis of the modified energy
loss rates are provided in Appendix B.

Graphically, the effects of the magnetic moment are shown in Panels
(b) and (c) of Fig.~\ref{second}. We see that the plasmon dominated
region enlarges considerably.  Comparing Fig.~\ref{second} with
Fig.~\ref{first}, we see that the $25\,\Msun$ star ``misses'' the plasmon
dominated region, while the $15\,\Msun$ and $10\,\Msun$  stars,
having lower central temperatures, enter it. Therefore, for them the
effects of the anomalous cooling should be important.

The above argument does not take into account possible feedback
effects of the extra cooling on the evolutionary trajectory. Indeed,
such feedback might be anticipated on general grounds: extra cooling,
if sufficiently strong, can lead to the decrease of temperature which
would in turn push the star deeper in the plasmon dominated degenerate
region.  Once the effects of degeneracy become important, cooling does
not change the pressure consistently and so there is not much
compressional heating as a reaction \citep{Arnett}. Thus, a
qualitative change in the evolutionary trajectory might be expected.

We will now see how these qualitative expectations are borne out by
the details numerical simulations.

\section{Massive Stars and non-standard Cooling: Results of Numerical Modeling}
\label{sect:numerical}

\subsection{Numerical modeling}

Our numerical modeling is based on KEPLER, a one-dimensional implicit
hydrodynamics stellar evolution code \citep{Weaver1978}. The physical
ingredients of this code -- such as the treatment of convection,
opacities, etc -- in the case of the standard physics are well
documented in the literature
\citep{Heger,WoosleyHeger2007,HegerWoosley2008}.

In order to simulate the effects of non-standard cooling in the
evolution of massive stars, we modified the neutrino cooling routine
of the code. Specifically, we computed the energy loss induced by the
electromagnetic pair production (see appendix) and derived an
interpolating function which is accurate at the level of $5\,\%$ in
all the region of interest.  The bremsstrahlung and photo processes
were left unchanged since they do not play any role in our analysis.%
\footnote{A simulation with the bremsstrahlung and photo induced
  cooling doubled showed, as expected, no sensitive changes in the
  evolution of the star.}
For the plasma process we used the fitting function given in
\citet{Raff3} which is shown to be better than $5\,\%$ in all the region
of interest for our purpose (see Fig.~8e of~\citealt{Raff3}).

We used the modified code to compute a two dimensional grid of data
points, varying the mass of the star, \mbox{$4\,\Msun<M<25\,\Msun$},
and the value of the neutrino magnetic moment,
\mbox{$0<\mu_\nu<0.5\times10^{-10}\,\muB$}.  The results are discussed
below.

\subsection{Results: Overview}

\begin{figure}
\centering
\includegraphics[angle=0, width=0.45\textwidth]{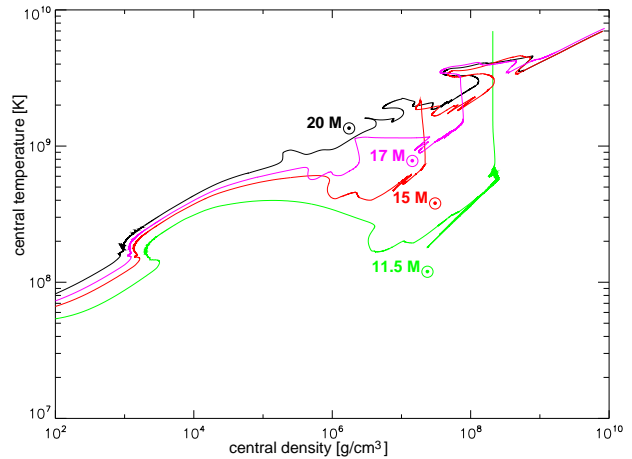}
\caption{The evolution of the central temperature and density with the
  extra cooling due to neutrino magnetic moment ($\mu_\nu=0.5\times
  10^{-10}\,\muB$). Four representative stellar masses ($11.5\,\Msun$,
  $15\,\Msun$, $17\,\Msun$, and $20\,\Msun$) were chosen. The
  $11.5\,\Msun$ star undergoes a thermonuclear CO explosion.}
 \label{core2}
\end{figure}

Figure~\ref{core2} displays the evolution of the central temperature
and density for stars of four representative masses: $11.5\,\Msun$,
$15\,\Msun$, $17\,\Msun$, and $20\,\Msun$.  These simulations include
our largest extra cooling of $\mu_\nu=0.5\times 10^{-10}\,\muB$.

As expected, we see that the evolution of the most massive of these
stars, $20\,\Msun$, shows no visible departure from the standard
trajectory (Fig.~\ref{first}). The lower mass stars, in contrast,
deviate from their standard trajectories into the region of lower
temperature. The curves start to bend between the He- and the
C-burning stages, that is as soon as the neutrino cooling becomes more
important than the radiation energy loss. The deviation becomes more
pronounced the lower the mass of the star, again as anticipated.

The physics behind this phenomenon is easy to understand. As the CO
core contracts following He burning, the heat that would otherwise
increase the temperature to the point of carbon ignition is instead
lost due to the additional cooling. The core is unable to stay on the
``ideal gas'' trajectory and instead collapses to a degenerate
configuration. The center of core is too cold to ignite carbon
burning, which is instead ignited \emph{off-center}. The center
becomes a relatively ``cold'' white dwarf ($T \sim (3-5) \times
10^8\,\K$) surrounded by a hotter burning shell ($T \sim 1 \times
10^9\,\K$).  Notice that while off-center ignition is observed in the
standard case for lower mass stars ($\sim 10\,\Msun$, see
Fig.~\ref{first}), here it is seen for stars of $15\,\Msun$.

What happens next depends on the mass of the star. In the case of the
$15\,\Msun$ star, the burning eventually reaches the center. At this
point, the corresponding curve in Fig.~\ref{core2} abruptly jumps back
up to the standard ``ideal gas'' trajectory. The silicon burning stage
in the center ensues and the star eventually undergoes iron core
collapse. The remarkable detour into the degenerate region results in
a somewhat different composition of the supernova progenitor (see
Sect.~\ref{sect:othereffects}).

A dramatically different fate awaits stars of lower mass, as
represented by the $11.5\,\Msun$ curve. In this case, the detour in
the degenerate region is more pronounced, with the central temperature
dropping down to $T \sim (1-2) \times 10^8\,\K$. The shell burning of
CO extinguishes before reaching the center.  As a result of the outer
shell burning, the relativistic degenerate CO core builds up until it
reaches the Chandrasekhar limit.  The core then undergoes a
thermonuclear runaway of carbon burning, consuming the CO fuel (CO
explosion).  This is indicated as indicated by the vertical line in
the Fig.~\ref{core2}.  In effect, the extra cooling leads to new kind
of supernova -- a SN Type Ia-like thermonuclear explosion inside a
massive star -- which perhaps could be viewed as new ``Type I.7''.
Our simulation stops at this point.

The fates of the stars of different masses as a function of the
neutrino magnetic moment are shown in Fig.~\ref{final}. We see that
neutrino magnetic moment appreciably shifts the mass thresholds
between the different outcomes.  In the presence of the extra cooling,
more and more massive stars are unable to ignite carbon burning.  The
CO window expands as a result and the ONeMg window shrinks.  The CO
explosion window first opens up at $\mu_\nu=0.2\times 10^{-10}\,\muB$
and for $\mu_\nu=0.5\times 10^{-10}\,\muB$ covers star with masses
$9.8\,\Msun\lesssim M\lesssim11.5\,\Msun$.  A complete overview of all
the models computed is given in \Tab{modov} of \Appendix{modov}.

\subsection{Changes of the WD/CCSN Thresholds}

\begin{figure}
\centering
\includegraphics[angle=0, width=0.45\textwidth]{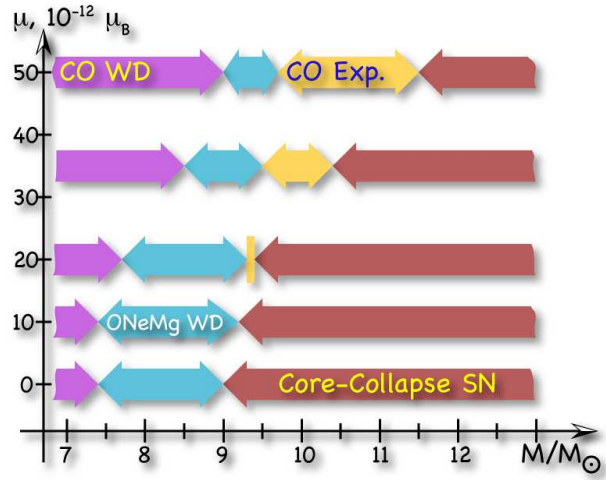}
\caption{The fate of the stars of different masses as a function of
  the neutrino magnetic moment (in units of $10^{-12}\,\muB$). The low
  mass stars end up as carbon-oxygen white dwarfs; heavier stars end
  up as oxygen-neon-magnesium white dwarfs (some of which result in
  electron capture supernovae); finally, still more massive stars
  undergo collapse in their iron core. As the magnetic moment is
  increased, the mass thresholds between the different types change
  and a new type -- a star undergoing a explosive thermonuclear
  runaway of its degenerate CO core -- appears.}
\label{final}
\end{figure}

As mentioned earlier, one of the effects of the extra cooling is to
make more difficult the ignition of various reactions, in particular
carbon, oxygen and Neon. To burn those nuclei now requires a larger
stellar mass, meaning that the threshold mass separating stars that
end up as white dwarfs from those that end up as supernovae increases,
as seen in Fig.~\ref{final}.

This may have interesting observational implications. In particular,
in recent years there has been important progress identifying
progenitors of observed supernovae. Hints are accumulating that stars
of initial mass $8\,\Msun-9\,\Msun$ do, in fact, end up as supernova
explosions. A particular case is supernova SN~2003gd (Type II-P), for
which the observations favor a red supergiant progenitor of mass
$\sim8\,\Msun-9\,\Msun$ \citep{VanDyk2003,Smartt2004,VanDyk2006}.

The situation with observations continues to improve at a rapid pace.
For example, very recently a progenitor of supernova SN~2008bk (also a
Type II-P) was identified to be a red supergiant with the mass of
$8.5\pm 1.0\,\Msun$ \citep{2008bk}. Likewise, a recent study of 20
progenitors of Type II-P supernovae finds that the minimum stellar
mass for these stars is $8_{-1.5}^{+1}\,\Msun$ \citep{smartt2008}.
Referring back to Fig.~\ref{final}, we see that values $\mu_\nu
\gtrsim 0.35 \times 10^{-10}\,\muB$ are already disfavored. It seems
highly likely that more observational progress is to follow.

The interpretation of the observations relies, among many factors, on
the stellar models used to infer the mass of the progenitor from its
color and luminosity. Hence, further refinement of the stellar models
would also be beneficial. To this end, it must be mentioned that the
magnetic moment cooling by itself does not change the relationship
between the surface characteristics and the mass: our simulation shows
changes in the central region of the star at late stages of the
evolution, but not in the surface characteristics.

\subsection{A New Type of Supernova}

The appearance of a new type of supernova as a consequence of the
additional cooling is perhaps the most remarkable finding of our
simulations. Should such objects have distinct characteristics not
observed in actual supernova explosions, a robust bound on the
magnetic moment, $\mu_\nu < 0.2 \times 10^{-10}\,\muB$, would result.

We again stress that we were unable to follow the development of the
explosion beyond the ignition of the degenerate CO burning.  Such
calculations would be highly desirable. Provisionally, we could
consider a plausible model of a Type Ia supernova exploding inside a
$\sim 9\,\Msun-11\,\Msun$ red supergiant. A simple estimate we
performed shows that, especially for stars in the low mass end of the
``CO explosions'' bin, the amount of energy available upon burning of
the entire CO core is enough to disrupt the star, though a detailed
hydrodynamic study to confirm this assessment has yet to be performed.
The resulting explosion would leave no neutron star remnant, and no
neutrino signal would be observed. The amount of $^{56}$Ni and
$^{56}$Co could be close to what is produced in a Type Ia explosion,
$\sim0.4\,\Msun-0.7\,\Msun$ \citep{Blinnikov2006}. This would
significantly exceed the amounts ejected in a standard core collapse
explosion, $\sim0.1\,\Msun-0.15\,\Msun$, or in possible ECSNe from our
ONeMg WD bin. As discussed by \citet{Heger}, the decay of $^{56}$Co to
$^{56}$Fe (half-life 77 days) is observationally very significant,
giving a ``radioactive tail'' to the light curve.  Thus, excessive
amounts of $^{56}$Co could be seen. For the current accuracy of
the determination of $^{56}$Ni mass from observations see, e.g., Fig.~9 in
\citet{smartt2008}.

\begin{figure}
\centering
\includegraphics[angle=0, width=\columnwidth]{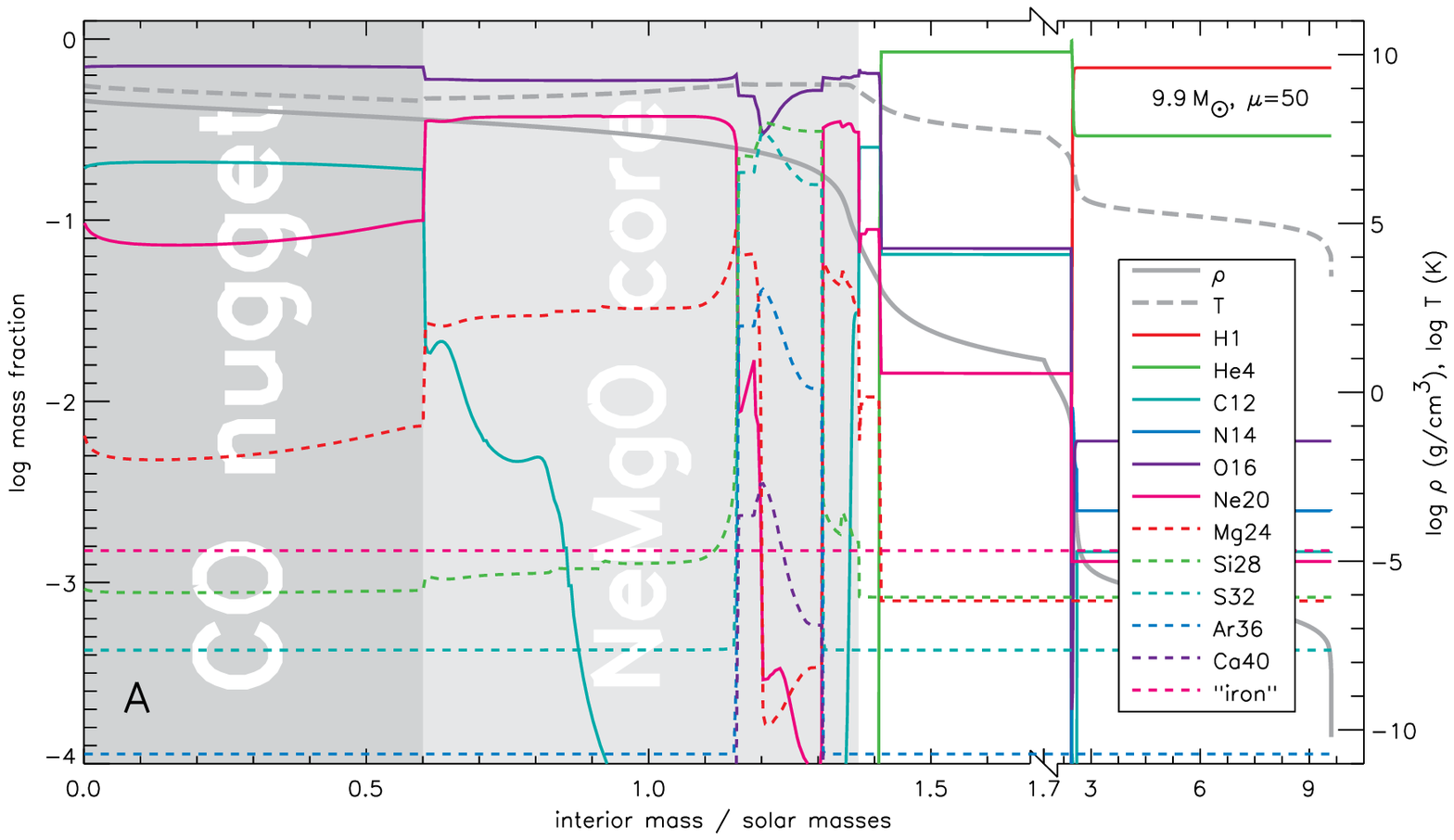}\\
\medskip
\includegraphics[angle=0, width=\columnwidth]{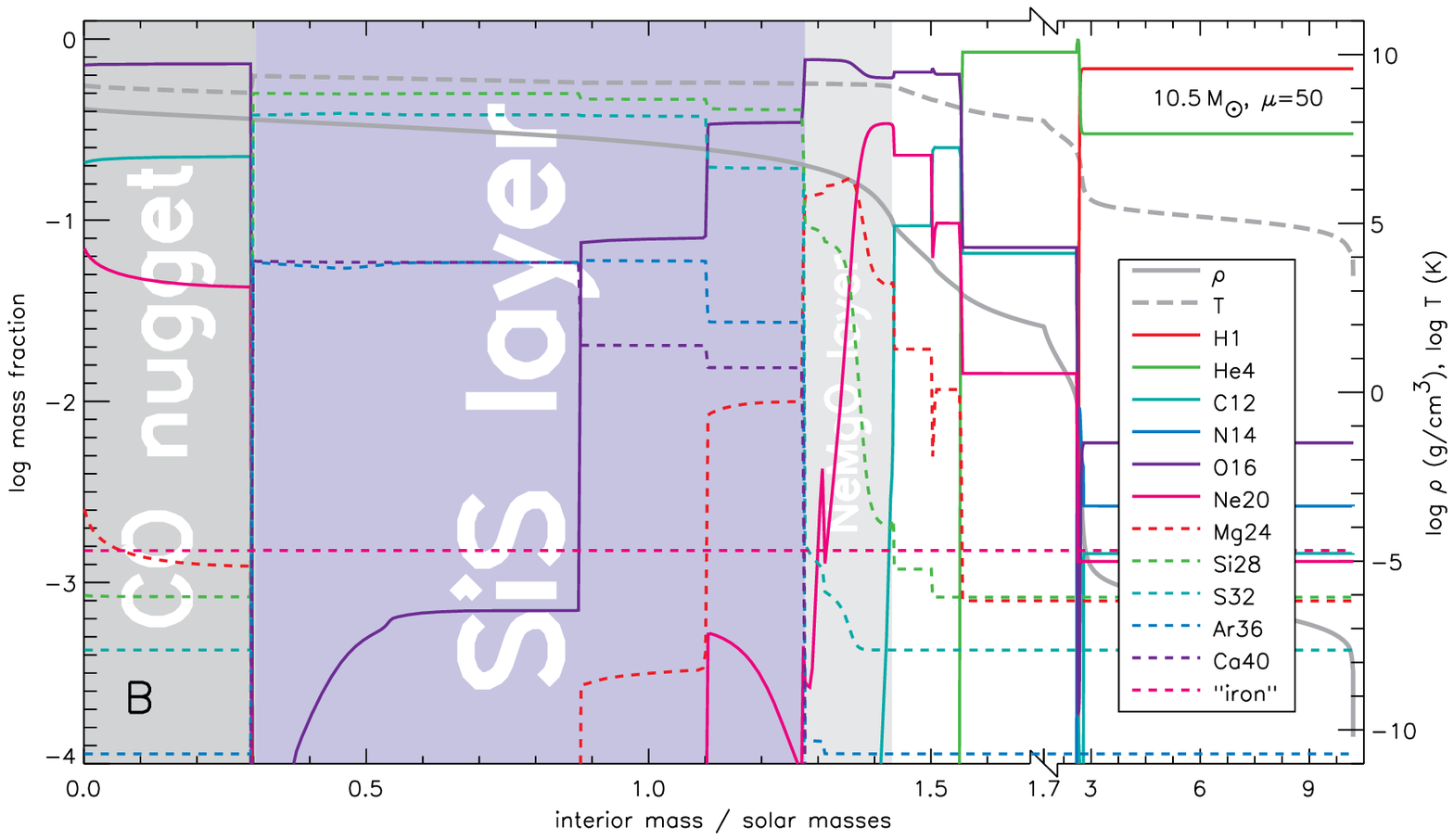}
\caption{Final structure of $9.9\,\Msun$ (Panel~A) and $10.5\,\Msun$
  (Panel~B) stars for $\mu_\nu=0.5\E{-10}\,\muB$ at the time of thermonuclear
  runaway at the center.
 \lFig{core_comp}}
\end{figure}

Notice that in the previous argument we concentrated on the low-mass
end of the ``CO explosions'' bin. In fact, the pre-explosion
composition of the stars in the ``CO explosions'' bin varies with the
mass of the star. Two typical examples are shown in \Fig{core_comp}.
We see that only the inner fraction of the core has a CO composition.
This CO inner core is surrounded by either a ONeMg layer at lower
masses, or a silicon-sulfur (SiS) layer (from ONeMg shell burning) at
higher masses.  We find that the size of the central CO
``nugget'' decreases as the mass of the star increases.  This will
affect densities, nucleosynthesis, and energetics of the explosion.

At the upper mass limit of the ``CO explosions'' bin, the
energy may eventually become insufficient to disrupt the star; in this
case, a burning pulse may eject the outer layer just before the final
core collapse SN, making it rather bright at the peak due to the
interaction of the core collapse SN ejecta and the previously ejected
layers. This mechanism is similar to the model for pair-instability
supernovae presented in \citet{WBH07}. The resulting SN would have
little \I{56}{Ni} to power the tail of the light curve.  

In a few cases at the upper mass limit we also observe the formation
of an ``iron'' layer from silicon shell burning; these cases have only
a tiny ($\sim0.1\,\Msun$) CO core left in the center and collapse as
core collapse SNe before igniting the central CO nugget; they would
probably be indistinguishable from regular CCSNe.

The detailed information from our simulations is collected in
\Tab{modov}. We can see that a significant fraction of the supernovae
in the ``CO explosions'' bin do have sizable CO cores, enough to
disrupt the stars and give the characteristic excess $^{56}$Co
observational signature.

Lastly, we note recent observations of the Kepler's supernova remnant
\citep{KeplerSNR}\footnote{We thank Carles Badenes for bringing this
  interesting study to our attention.}. The abundance of iron in the
shocked ejecta together with no evidence for oxygen-rich ejecta indicate
that this object underwent a thermonuclear explosion. Curiously, a
detailed analysis described in that paper suggests the possibility of
a Type Ia explosion in a more massive progenitor. That the explosion
occurred in the environment that is not low in metallicity is
especially interesting. Clearly, further observational studies are
very desirable.

\subsection{Other Effects}
\label{sect:othereffects}

There are several other effects of the extra cooling that are worth
mentioning. For example, the resulting composition of the
core-collapse supernova progenitor is somewhat modified.  Our
calculations of the $12\,\Msun$ star show the size of the CO core
decreases by $8\,\%$ with $\mu_\nu=0.5\times 10^{-10}\,\muB$. This small
change could be of interest, since the explosion mechanism is a
sensitive function of the chemical composition.

Another effect is the change of the presupernova neutrino signal
\citep{presn}. This occurs because the last stage before the iron core
collapse -- Si burning -- proceeds significantly faster since the
energy escapes the core more easily.  For $12\,\Msun$ and
$\mu_\nu=0.35\times 10^{-10}\,\muB$, the Si shell burning time is only
about 8 hours, instead of 15 hours for $\mu_\nu=0$.  The pre-supernova
neutrino signal at a future megaton-sized neutrino detector could be
shorter with higher luminosity.

Other effects may be revealed upon a more detailed investigation. For
example, the extra cooling would modify the nucleosynthetic yields of
various elements.

\section{Conclusion and Perspectives}
\label{sect:conclusions}

For certain kinds of new physics, useful astrophysical constraints can
be obtained using simple order-of-magnitude analytical estimates.
Examples are provided by various theories with extra dimensions (see,
e.g.,
\citealt{CullenPerelstein,HannestadRaffelt,FriedlandGiannotti}).  In
contrast, the treatment of the problem given here requires a detailed
analysis with a sophisticated stellar evolution code.

As we have seen, stars with masses in the range $7\, \Msun
\lesssim M \lesssim 18\, \Msun$ are indeed sensitive to the neutrino
magnetic moment. The additional cooling it introduces does not simply
change the evolutionary time scales of these stars, but leads to the
appearance of \emph{qualitatively new features} in the evolution. The
physics involved is very rich and more detailed investigations of many
aspects are clearly desirable. We have undertaken here the first step
to explore the subject.  Let us now summarize our main results and
consider some perspectives.

\emph{1)} We have shown that the magnetic moment of the size $\mn \sim
5 \times 10^{-11}\,\muB$, which is below the experimental bound, would
modify the evolution of stars with the masses $\lesssim 17\,\Msun$. The
evolution of heavier stars is unchanged. This is a consequence of the
fact that heavier stars have higher central temperatures and as a
result ``miss'' the region of the parameter space where the extra
cooling operates.

\emph{2)} The minimal value of $\mn$ to which the star is sensitive
depends on the initial mass.  For stars whose initial mass is about
$9-10\, \Msun$, this value is as low as $2 \times 10^{-11}\,\muB$.  The
sensitivity of massive stars to the neutrino magnetic moment is thus
comparable to that of the HB stars \citep{RaffHB}. Notice that
\citet{Raff1} found that the analysis of RG cooling yields a more
stringent bound, $\mn\lesssim3 \times 10^{-12}\,\muB$.

\emph{3)} The extra cooling due to the neutrino magnetic moment shifts
the threshold masses for the different final stages of the evolution.
The number of carbon-oxygen white dwarfs increases, while the
corresponding number of the oxygen-neon-magnesium white dwarfs and
core collapse supernovae decreases. These shifts are graphically
illustrated in Fig~\ref{final}. This effect could be constrained using
observations of supernova progenitors. Such data is now coming in at a
rapid pace.

\emph{4)} According to our code, stars with initial mass $\sim9.3\,
\Msun$ undergo a thermonuclear explosion for values of $\mn$ larger
than $2 \times 10^{-11}\,\muB$.  This is a new feature, absent in the
standard model evolution, and in principle observable.  The initial
mass region for which this happens enlarges with larger $\mn$, and is
depicted in Fig.~\ref{final}, labeled as ``CO explosion''.  For $\mn=
5 \times 10^{-11}\,\muB$ we predict that roughly one in five of the
stars which do not become WDs end their lives with this thermonuclear
explosion, whereas the others become CCSNe.

\emph{5)}
The final composition of the presupernova is a function of $\mn$.
This might be of interest since the explosion mechanism is  a sensitive
function of the chemical composition.

\emph{6)} For stars that end their lives as CCSNe, a possible
presupernova neutrino signal at a future megaton water Cherenkov
neutrino detector would be modified in the presence of the
additional cooling. The neutrino flux would be higher and the duration
shorter.

\emph{7)} Our conclusions are based on a numerical simulation performed
with the KEPLER code.  
A reliable prediction of the threshold masses between different kinds
of WDs and CCSN depends on, among other ingredients, the
treatment of convection, which varies between the codes.
A comparison of different codes (standard evolution only) was
performed by \citet{Heger2}, yielding slightly different predictions
of the thresholds.  In the case of nonstandard physics, a similar
comparison would be very useful.  Additionally, one has to keep in
mind other uncertainties that affect stellar evolution, such as the
initial stellar composition (e.g., \citealt{THA07}) and the natural
distribution of stellar rotation rates (e.g., \citealt{HLW00}).

\emph{8)} 
Beyond this list, various other aspects of the extra cooling
could be considered, such as the
changes in the nucleosynthesis of various elements.

\acknowledgements 

This work was performed under the auspices of the National Nuclear
Security Administration of the U.S. Department of Energy at Los Alamos
National Laboratory under Contract No.  DE-AC52-06NA25396.  We thank
C.~Badenes, C.~Fryer, J.~Beacom, and S.-C.~Yoon for valuable
suggestions and discussions.  AH has also been supported, in part, by
the DOE Program for Scientific Discovery through Advanced Computing
(SciDAC; DOE-FC02-01ER41176 and DOE-FC02-06ER41438) and by the US
Department of Energy under grant DE-FG02-87ER40328.

\appendix

\section{APPENDIX A:  Energy loss due to standard neutrino interactions}

In this appendix we give approximate expressions for the \emph{standard}  neutrino 
cooling rates as a function of density and temperature, in order to give 
an analytical description of Fig.~\ref{second}a. 
We restrict ourselves only to the regions relevant to the evolution of massive stars.   

At temperatures below $\lesssim 5 \times 10^{8}\,\K$ and low
densities (orange region in the middle left in Fig.~\ref{second}a)
the photo process dominates. Energy loss per unit mass due to this
process is roughly \citep{Salpeter2}
\begin{equation} \label{photo}
        \ePh\simeq 0.1\, \erggs \times T_8^8\,,
\end{equation}
where $T_8=T/10^8\,\K$.
At higher temperatures (top turquoise region in Fig.~\ref{second}a)
pair production instead dominates. The energy loss is roughly\citep{Fowler} 
\begin{equation}\label{eq:pairlowT}
\eP \simeq 4.6\E{11}\, \erggs
        \, \frac{T_8^3}{\rho_4}\,
        e^{-118.5/T_8} \,,
        \end{equation}
for $T<{\rm few}~ 10^9\,\K$ and  
\begin{equation}\label{eq:pairhighT}
\eP\simeq 2.8\E{20}\, \erggs
        \, \frac{T_{10}^9}{\rho_4} \,,
\end{equation}
for $T>{\rm few}~ 10^9\,\K$,
where $\rho_4=\rho/10^4\,\gcc$ and $T_{10}=T/10^{10}\,\K$.

Both the above equations apply only for non-degenerate plasma.  When
degeneracy becomes important the pair emission is exponentially
suppressed by the electron chemical potential.  In this regime, the
main loss mechanism is the neutrino emission via plasmon decay (green
region in the middle of Fig.~\ref{second}a). This rate was first
calculated by \citet{Inman}.  For densities such that $\opl \lesssim
T$, the energy loss due to this process is approximately
\begin{equation}
        \ePl\simeq 0.0127\, \erggs \, T_8^3\, \rho_4^2.
\end{equation}
At very high densities ($\opl \gg T$), this emission is exponentially
suppressed by the plasma frequency and the bremsstrahlung process
becomes the dominant one.

\section{APPENDIX B:  Energy loss due to non-standard neutrino  interactions}

If neutrinos have a non-vanishing magnetic moment, the production
mechanisms (pair, plasma, photon, bremsstrahlung) would receive an
\emph{additional} electromagnetic contribution from the interaction
term of Eq.(\ref{Lint}), as shown in Fig.~\ref{diagramMu}.  We
restrict ourselves only to the processes relevant for the evolution of
massive stars, namely plasma and pair production (the photo process
dominates neutrino energy loss in a region where the star cools mainly
via radiative energy loss).
 
\subsection{Plasma processes} 

The standard and electromagnetically induced plasmon emission rates 
($\ePl $  and $\ePl^\mu$)  depend  differently on the plasma 
frequency, and hence on the density. 
This behavior can be anticipated on the grounds of dimensional
analysis. 
The energy loss rate can be schematically written as $(\mbox{decay
  rate})\times (\mbox{photon energy}) \times (\mbox{number density of
  photons})$.  Therefore the ratio of the standard $\ePl$ and the
additional nonstandard $\ePl^\mu$ cooling rates is equal to the
corresponding ratio of the plasmon decay rates, $\Gamma^\mu/\Gamma$
({cf.  \citealt{FriedlandGiannotti}}).  The latter, in turn, can be
estimated by dimensional analysis. The rates are proportional to the
squares of the corresponding coupling constants, $\mn$ or $\GF/e$.
The remaining dimension should be fixed by $\opl$ alone. (In the rest
frame of the plasmon, $T$ cannot enter.) Therefore, 
$\Gamma^{\mu}/\Gamma \sim \mn^2/[G_F^2/\alpha] \opl^{-2}$.  In the
region of temperature and density of interest for us, a detailed
calculation gives \citep{Raff3}
\begin{equation}
\label{plasma}
\frac{\ePl^\mu}{\ePl}\simeq 0.318 \, \left( \frac{10\,\keV}{\opl} \right)^2 \m^2,
\end{equation}
where the plasma frequency is numerically given by, {\it e.g.},
\citet{RaffBook},
\begin{equation}\label{opl}
\opl=28.7\,\eV\frac{(\Ye \rho)^{1/2}}{[1+(1.019\times 10^{-6}\,\Ye \rho)^{2/3}]^{1/4}},
\end{equation}
with $\rho$ in $\gcc$.  Therefore at low densities, $\rho\ll 10^6
\gcc$, the ratio between the electromagnetic induced plasmon emission
rate and the standard one is inversely proportional to the density,
whereas at high density $\rho\gg 10^6 \gcc$, it is inversely
proportional to $\rho^{2/3}$.  Finally, taking for example $\mn/\muB =
10^{-10}$, equation (\ref{plasma}) shows that, for density
sufficiently low, the non-standard plasma process can be much larger
than the standard one.  Therefore, when the non-standard neutrino
interactions are turned on, the region where the plasma process is
dominant becomes larger, especially at low densities (see panel (b)
and (c) of Fig.~\ref{second}).

\subsection{Pair production}

The energy loss per unit time and volume in neutrino pair production
is given by the transition probability for $e^+ e^- \to \nu \bar{\nu}$
annihilation multiplied by the neutrino pair energy and integrated
over the density of both initial and final states.  This reads
\begin{eqnarray}
        \QP=\frac{4}{(2 \pi)^6}\int\frac{d^3p_1}{e^{(E_1-\mu)/T}+1}\frac{d^3p_2}{e^{(E_2
        +\mu)/T}+1}  (E_1+E_2)    \vrel\,\sP~,
\end{eqnarray}
where $\mu$ is  the chemical potential, $\vrel$ is the electron-positron relative velocity  and
 $\sP$ represents the spin averaged  cross-section for $e^+  e^- \to \nu \bar{\nu}$ annihilation.
The quantity $E_1 E_2 \vrel \sP$ is Lorentz invariant and is given by
($p_i = (E_i, {\bf p}_i)$, $q_i = (\omega_i, {\bf q}_i)$)
\begin{eqnarray}
E_1 E_2 \vrel \, \sP=\frac{1}{4} \int
\frac{d^3q_1\,d^3q_2}{(2\pi)^3 2\omega_1\,(2\pi)^3  2 \omega_2}  |M|^2  \, (2\pi)^4\,\delta^{4}(p_1+p_2-q_1-q_2) \ \ \ ~
\label{eq:vrsigma}
\end{eqnarray}
in terms of the spin-averaged squared invariant amplitude $|M|^2$.
If the pair production proceeds via neutrino  magnetic moment,
we find   ($s= (p_1 + p_2)^2$, $t= (p_1 - q_1)^2$, $u= (p_1 - q_2)^2$)
\begin{equation}
|M_{\mu_\nu}|^2= \frac{e^2\mu_\nu^2}{s} \left[  s^2 -  (u - t)^2 \right] ~,
\end{equation}
where $\mu_\nu^2$ represents the effective dipole moment:
\begin{equation}
\mu_\nu^2 = \sum_{i,j=1}^{3}  \ |\mu_{ij} |^2~.
\end{equation}
Integrating over the invariant phase space leads to
\begin{equation}
        E_1 E_2 \vrel \sP_{\mu_\nu}=
         \frac{\alpha \, \mu_\nu^2}{12}  \left(s+2\me^2\right)~,
\end{equation}
or equivalently $\vrel \sP_{\mu _\nu} =  (\alpha \mu_\nu)^2/3 \, (1 + 2 \me^2/s)$.
Performing the angular integrations in Eq.~\ref{eq:vrsigma}  we then find
\begin{eqnarray}
        \Qu=\frac{\alpha \, \mu_\nu^2}{6 \pi^4}\int_{\me}^{\infty} \, dE_1\; dE_2
        \ p_1 \, p_2  \frac{(E_1+E_2)(2\me^2+E_1 E_2)}{(e^{(E_1-\mu)/T}+1)(e^{(E_2+\mu)/T}+1)}~,
\label{eq:qpair1}
\end{eqnarray}
showing that the expression for the luminosity factorizes in the product of two one-dimensional
integrals.

Following \citet{Salpeter}, we now define the dimensionless variables
\begin{equation}
\lambda=T/\me    \ ,\qquad \nu=\mu/T~,
\end{equation}
and the dimensionless integrals 
\begin{equation}
\label{G}
        G^{\pm}_{n}(\lambda,\nu)= \lambda^{3+2n}
        \int_{\lambda^{-1}}^{\infty}
        dx \frac{x^{2n+1}(x^2-\lambda^{-2})^{1/2}}{e^{(x\pm\nu)}+1}~.
\end{equation}
The luminosity  due to non-standard pair production reads
\begin{eqnarray}
        \Qu=\frac{\alpha \, \mu_\nu^2  \,\me^7}{6 \pi^4}
        \Big[ 2G^-_{-1/2}G^+_{0}+2G^-_{0}G^+_{-1/2}+G^-_{0}G^+_{1/2}+G^-_{1/2}G^+_{0}\Big]~,
\label{eq:pairmu}
\end{eqnarray}
and is a function of the temperature and the chemical potential.
In order to have a function of temperature and density, the chemical potential must be
solved in terms of temperature and density. This requires the solution of the following 
equation\citep{Salpeter}
\begin{equation}
\label{rho}
        \rho= 2.922\times 10^6\,\gcc\, (G^{-}_{0}-G^{+}_{0})  \,, 
\end{equation}
which cannot in general be inverted analytically in terms of $\mu$.
The numerical solutions are plotted in Fig.~1 of \citet{Salpeter}.

\begin{figure}
\centering
\includegraphics[angle=0, width=0.45\textwidth]{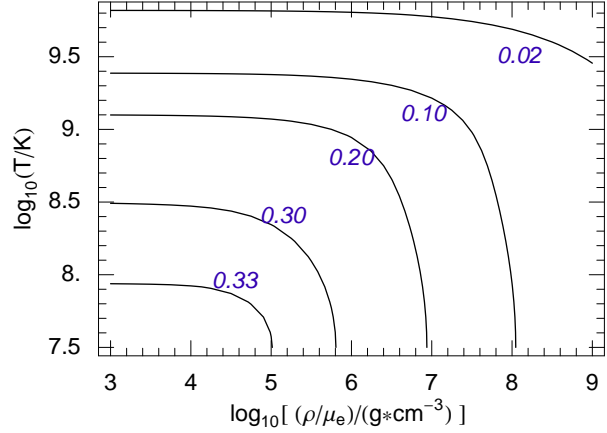}
\caption{Contours of the ratio $\QP_{\mn}/\Qsm$ of
  non-standard over standard energy loss via neutrino pair production
  as a function of temperature and density.  The effective neutrino
  magnetic moment is fixed at $\mu_{10} = 1$.}
 \label{fig:Qpair}
\end{figure}

In order to assess the impact of non-standard cooling,
the above result has to be compared to the  Standard Model emission rate~\citep{Dicus}:
\begin{eqnarray}
\Qsm (\nue) &=&\frac{\GF^2 \me^9}{18\pi^5}\Big\{
(7 \CV^2-2\CA^2)[G^-_{0}G^+_{-1/2}+G^-_{-1/2}G^+_{0}]
+9\CV^2[G^-_{1/2}G^+_{0}+G^-_{0}G^+_{1/2}]    \nonumber \\
&+&(\CV^2+\CA^2)
[4G^-_{1}G^+_{1/2}+4G^-_{1/2}G^+_{1}
-G^-_{1}G^+_{-1/2}
-G^-_{1/2}G^+_{0}-G^-_{0}G^+_{1/2}-G^-_{-1/2}G^+_{1}]
\Big\}~.
\label{eq:pairSM}
\end{eqnarray}
Here $\CV= 1/2 + 2 \sin^2 \tW$ and $\CA = 1/2$, with the weak mixing angle
numerically given by $\sin^2 \tW = 0.23$.
The energy loss due to $\nu_{\mu}$ and $\nu_\tau$ is obtained form
$\Qsm (\nue)$ by replacing  $\CV \to \CV-1$ and $\CA \to - \CA$.

A simple comparison of the standard and non-standard pair production
amplitudes suggests that the two contributions are of comparable size
if $\mu_\nu \sim 10^{-10} \,\muB$.  Using the full expressions of
Eqs.~\ref{eq:pairmu} and \ref{eq:pairSM} we plot in
Fig.~\ref{fig:Qpair} the ratio $\Qu/\Qsm $ for $\mu_\nu = 10^{-10}
\,\muB$ in the region of interest in the $\rho-T$ plane.
The results show that indeed  the non-standard loss never exceeds $30\% $ of the electro-weak one,
being most important in the non-relativistic and non-degenerate region. 
From the behavior of  $G_{n}^{\pm} (\lambda, \nu)$ in the relativistic and/or
degenerate limit one can simply verify that $\Qu/\Qsm \to 0$
in these physical regimes, as shown in Fig.~\ref{fig:Qpair}.

For the numerical implementation in the stellar evolution code we have used the simple
fitting formula
\begin{eqnarray}
\frac{\Qu}{\Qsm} (\bar{\rho}, \bar{T})  &=&  \
 \mu_{10}^2  \times \Big[ n(\bar{\rho}) +  a(\bar{\rho}) (\log_{10} \bar{T} - 7.5)^2
 +    b(\bar{\rho}) (\log_{10} \bar{T} - 7.5)^{10}   \Big]^{-1},
\\
n(\bar{\rho}) &=&   2.959105 + 0.0021515
(\log_{10}\bar{\rho} - 3)^5,
\\
a(\bar{\rho}) &=&     0.362925  + 0.000408304  \, \bar{\rho}^{0.436811},
\\
b(\bar{\rho}) &=&  0.00998357 +  2.76174 \cdot 10^{-7}  \,   \bar{\rho}^{0.5018889}~,
\end{eqnarray}
where $\mu_{10}$ is defied by $\mu_{\nu} = \mu_{10} \, 10^{-10} \,\muB$,
$\bar{T} = T/\K$, and $\bar{\rho} = (\rho/\mu_e)/(\gcc)$.
The fitting formula is accurate at the level of  $5\,\% $  or better in the region defined by
$10^{7.5} < \bar{T} < 10^{10}$ and
$10^{3} < \bar{\rho} < 10^{10}$.

Finally, in the region where the pair process dominates neutrino
emission, the additional energy loss per unit time and mass is approximately
given by:
\begin{equation}
       \eP^\mu\simeq 1.6 \times 10^{11}\,  \erggs   \times 
       \frac{ \mu_{10}^2}{ 
       \rho_4}\,
       e^{-118.5/T_8}\,    
        ~, 
\end{equation}
where $\mu_{10}=\mn/10^{10}\,\muB$.
Comparing with Eqs.~(\ref{eq:pairlowT}) and (\ref{eq:pairhighT}) we see that this energy loss is never more that $\sim 30\%$ of the standard model value, if $\mn$ is below the experimental bound.

\section{APPENDIX C:  Detailed Listing of Model Results}
\lSect{modov}

\Tab{modov} summarizes our calculations and results for different values of $\mn$. 

\begin{table}
  \caption{Overview of Model Results\lTab{modov}}
\centering
\begin{tabular}{rl}
\hline\hline
\noalign{\smallskip}
$M/\Msun$ & Comment\\
\noalign{\smallskip}
\hline
\noalign{\smallskip}
\multicolumn{2}{c}{$\mn=0.00\E{-10}\,\muB$}\\
\noalign{\smallskip}
\hline
\noalign{\smallskip}
$4.0-7.1 $ &   AGB, CO WD \\
$7.2 $ &   AGB, CO WD with $0.40\,\Msun$ NeMgO outer layer, $0.70\,\Msun$ CO center \\
$7.3 $ &   AGB, ONeMg WD, 3rd dredge-up, vast C shell burning, $0.02\,\Msun$ CO core left in center \\
$7.4-8.9 $ &   AGB, OMeMg WD, 3rd dredge-up \\
$9.0-9.1 $ &   3rd dredge-up, may make ECSN \\
$9.2 $ &   NeO thermonuclear runaway, weak explosion; fallback could still be possible \\
$9.3 $ &   CCSN, $0.21\,\Msun$ ONeMg core, explosive ignition inside Fe shell \\
$9.4-25.0 $ &   CCSN \\

\noalign{\smallskip}
\hline
\noalign{\smallskip}
\multicolumn{2}{c}{$\mn=0.10\E{-10}\,\muB$}\\
\noalign{\smallskip}
\hline
\noalign{\smallskip}

$4.0-7.4$&AGB, CO WD \\
$7.5-8.9$&AGB, ONeMg WD, 3rd dredge-up, vast C shell burning, up to $0.18\,\Msun$ CO core left in center \\
$9-9.1$&AGB, ONeMg WD (3rd dredge-up) \\
$9.2  $&ONeMg WD or ECSN  \\
$9.3  $&CCSN, ONeMg core with SI Shell burning,  \\
$9.4  $&CCSN, central NeO ignition, weak explosion, fallback of Fe core \\
$9.5-25.0  $&CCSN \\
       
\noalign{\smallskip}
\hline
\noalign{\smallskip}
\multicolumn{2}{c}{$\mn=0.20\E{-10}\,\muB$}\\
\noalign{\smallskip}
\hline
\noalign{\smallskip}

$4.0-7.7$ & AGB, CO WD \\
$7.8-9.2$ & AGB, ONeMg WD \\
$9.3$ &  O/Ne/C thermonuclear runaway \\
$9.4$ & $0.32\,\Msun$ SiS core ($10\,\%$ O), CCSN \\
$9.5$ &  Si explosion, 3rd dredge-up of He shell, CCSN \\
$9.6, 9.8$ & CCSN, Si detonation\\
$9.7, 9.9 $ & CCSN\\
$10.0$ &  CCSN, $0.20\,\Msun$ SiS nugget at center, should make Si detonation\\
$10.5-25.0$ & CCSN\\

\noalign{\smallskip}
\hline
\noalign{\smallskip}
\multicolumn{2}{c}{$\mn=0.35\E{-10}\,\muB$}\\
\noalign{\smallskip}
\hline
\noalign{\smallskip}

$4.0-8.0$ & AGB, CO WD\\
$8.1-8.4$ & AGB, $1.08-1.13\,\Msun$ CO WD with $0.10-0.40\,\Msun$ ONeMg layer \\
$8.5-9.4$ & AGB, $1.15-1.32\,\Msun$ ONeMg WD with $\sim25\,\%$ C in center \\
$9.5$ & AGB, $1.35\,\Msun$ ONeMg, $25\,\%$ C in center, deep 3rd dredge-up \\
$9.6-9.9$   & C thermonuclear runaway, $0.63-0.50\,\Msun$ CO nugget, inside $1.37-1.00\,\Msun$ ONeMg core \\
$10.0-10.4$ & C thermonuclear runaway, $0.25-0.03\,\Msun$ CO nugget in center \\
$10.5-25.0$ & CCSN\\ 

\noalign{\smallskip}
\hline
\noalign{\smallskip}
\multicolumn{2}{c}{$\mn=0.50\E{-10}\,\muB$}\\
\noalign{\smallskip}
\hline
\noalign{\smallskip}

$4.0$     & AGB, CO WD, $0.2\,\%$ He left close to center\\
$4.5-7.5$ & AGB, CO WD \\
$8.0-8.3$ & AGB, CO WD, starting and increasing C shell burning\\
$8.4-9.0$ & AGB, CO WD, $0.20-0.60\,\Msun$ layer with Ne $>$ C\\
$9.1-9.6$ & AGB, ONeMg WD, $0.60-0.49\,\Msun$ CO core\\
$9.7$ &  AGB, ONeMg WD, $0.52\,\Msun$ CO core, no 3rd dredge-up\\
$9.8$ &  C thermonuclear runaway, $0.7\,\Msun$ CO core, $1.36\,\Msun$ He core, 3rd dredge-up, NeO shell\\
$9.9-10.4$ & C thermonuclear runaway, $0.60-0.49\,\Msun$ CO core, $2.46-2.62\,\Msun$ He core,  NeO shell\\
$10.5-10.9$ & C thermonuclear runaway, $0.30-0.10\,\Msun$ CO core, $2.66-2.78\,\Msun$ He core,  SiS shell\\
$11.0$ &   C thermonuclear runaway, $0.02\,\Msun$ CO core, $2.810\,\Msun$ He core, Fe shell $\Rightarrow$ CCSN\\ 
$11.1$ & C thermonuclear runaway, $0.05\,\Msun$ CO core, $2.80\,\Msun$ He core,  SiS shell\\
$11.2$ &   C thermonuclear runaway, $0.05\,\Msun$ CO core, $2.82\,\Msun$ He core, Fe shell $\Rightarrow$ CCSN\\ 
$11.3$ & C thermonuclear runaway, $0.17\,\Msun$ CO core, $2.76\,\Msun$ He core,  SiS shell\\
$11.4$ & C thermonuclear runaway, $0.35\,\Msun$ CO core, $2.62\,\Msun$ He core,  NeO shell\\
$11.5$ & C thermonuclear runaway, $0.15\,\Msun$ CO core, $2.74\,\Msun$ He core,  SiS shell\\
$11.6-25.0$ &  CCSN\\

\noalign{\smallskip}
\hline
\end{tabular}
\end{table}

\end{document}